\documentclass[twocolumn,amsmath,10pt,floatfix,prl,aps,superscriptaddress]{revtex4-1}

\usepackage[T1]{fontenc}
\usepackage[latin9]{inputenc}
\usepackage{graphicx}

\begin{document}
%% STAR Collaboration authors and affiliations (Oct.18, 2018)
\affiliation{Abilene Christian University, Abilene, Texas   79699}
\affiliation{AGH University of Science and Technology, FPACS, Cracow 30-059, Poland}
\affiliation{Alikhanov Institute for Theoretical and Experimental Physics, Moscow 117218, Russia}
\affiliation{Argonne National Laboratory, Argonne, Illinois 60439}
\affiliation{Brookhaven National Laboratory, Upton, New York 11973}
\affiliation{University of California, Berkeley, California 94720}
\affiliation{University of California, Davis, California 95616}
\affiliation{University of California, Los Angeles, California 90095}
\affiliation{University of California, Riverside, California 92521}
\affiliation{Central China Normal University, Wuhan, Hubei 430079 }
\affiliation{University of Illinois at Chicago, Chicago, Illinois 60607}
\affiliation{Creighton University, Omaha, Nebraska 68178}
\affiliation{Czech Technical University in Prague, FNSPE, Prague 115 19, Czech Republic}
\affiliation{Technische Universit\"at Darmstadt, Darmstadt 64289, Germany}
\affiliation{E\"otv\"os Lor\'and University, Budapest, Hungary H-1117}
\affiliation{Frankfurt Institute for Advanced Studies FIAS, Frankfurt 60438, Germany}
\affiliation{Fudan University, Shanghai, 200433 }
\affiliation{University of Heidelberg, Heidelberg 69120, Germany }
\affiliation{University of Houston, Houston, Texas 77204}
\affiliation{Indiana University, Bloomington, Indiana 47408}
\affiliation{Institute of Modern Physics, Chinese Academy of Sciences, Lanzhou, Gansu 730000 }
\affiliation{Institute of Physics, Bhubaneswar 751005, India}
\affiliation{University of Jammu, Jammu 180001, India}
\affiliation{Joint Institute for Nuclear Research, Dubna 141 980, Russia}
\affiliation{Kent State University, Kent, Ohio 44242}
\affiliation{University of Kentucky, Lexington, Kentucky 40506-0055}
\affiliation{Lawrence Berkeley National Laboratory, Berkeley, California 94720}
\affiliation{Lehigh University, Bethlehem, Pennsylvania 18015}
\affiliation{Max-Planck-Institut f\"ur Physik, Munich 80805, Germany}
\affiliation{Michigan State University, East Lansing, Michigan 48824}
\affiliation{National Research Nuclear University MEPhI, Moscow 115409, Russia}
\affiliation{National Institute of Science Education and Research, HBNI, Jatni 752050, India}
\affiliation{National Cheng Kung University, Tainan 70101 }
\affiliation{Nuclear Physics Institute of the CAS, Rez 250 68, Czech Republic}
\affiliation{Ohio State University, Columbus, Ohio 43210}
\affiliation{Institute of Nuclear Physics PAN, Cracow 31-342, Poland}
\affiliation{Panjab University, Chandigarh 160014, India}
\affiliation{Pennsylvania State University, University Park, Pennsylvania 16802}
\affiliation{Institute of High Energy Physics, Protvino 142281, Russia}
\affiliation{Purdue University, West Lafayette, Indiana 47907}
\affiliation{Pusan National University, Pusan 46241, Korea}
\affiliation{Rice University, Houston, Texas 77251}
\affiliation{Rutgers University, Piscataway, New Jersey 08854}
\affiliation{Universidade de S\~ao Paulo, S\~ao Paulo, Brazil 05314-970}
\affiliation{University of Science and Technology of China, Hefei, Anhui 230026}
\affiliation{Shandong University, Qingdao, Shandong 266237}
\affiliation{Shanghai Institute of Applied Physics, Chinese Academy of Sciences, Shanghai 201800}
\affiliation{Southern Connecticut State University, New Haven, Connecticut 06515}
\affiliation{State University of New York, Stony Brook, New York 11794}
\affiliation{Temple University, Philadelphia, Pennsylvania 19122}
\affiliation{Texas A\&M University, College Station, Texas 77843}
\affiliation{University of Texas, Austin, Texas 78712}
\affiliation{Tsinghua University, Beijing 100084}
\affiliation{University of Tsukuba, Tsukuba, Ibaraki 305-8571, Japan}
\affiliation{United States Naval Academy, Annapolis, Maryland 21402}
\affiliation{Valparaiso University, Valparaiso, Indiana 46383}
\affiliation{Variable Energy Cyclotron Centre, Kolkata 700064, India}
\affiliation{Warsaw University of Technology, Warsaw 00-661, Poland}
\affiliation{Wayne State University, Detroit, Michigan 48201}
\affiliation{Yale University, New Haven, Connecticut 06520}

\author{J.~Adam}\affiliation{Creighton University, Omaha, Nebraska 68178}
\author{L.~Adamczyk}\affiliation{AGH University of Science and Technology, FPACS, Cracow 30-059, Poland}
\author{J.~R.~Adams}\affiliation{Ohio State University, Columbus, Ohio 43210}
\author{J.~K.~Adkins}\affiliation{University of Kentucky, Lexington, Kentucky 40506-0055}
\author{G.~Agakishiev}\affiliation{Joint Institute for Nuclear Research, Dubna 141 980, Russia}
\author{M.~M.~Aggarwal}\affiliation{Panjab University, Chandigarh 160014, India}
\author{Z.~Ahammed}\affiliation{Variable Energy Cyclotron Centre, Kolkata 700064, India}
\author{I.~Alekseev}\affiliation{Alikhanov Institute for Theoretical and Experimental Physics, Moscow 117218, Russia}\affiliation{National Research Nuclear University MEPhI, Moscow 115409, Russia}
\author{D.~M.~Anderson}\affiliation{Texas A\&M University, College Station, Texas 77843}
\author{R.~Aoyama}\affiliation{University of Tsukuba, Tsukuba, Ibaraki 305-8571, Japan}
% ...

\collaboration{STAR Collaboration}\noaffiliation

\title{Measurements of Dielectron Production in Au$+$Au Collisions
at $\sqrt{s_{NN}}$= 27, 39, and 62.4~GeV from the STAR Experiment}

\begin{abstract}
We report systematic measurements of dielectron ($e^{\pm}e^{\pm}$) invariant-mass $M_{ee}$ 
spectra at mid-rapidity in Au+Au collisions at $\sqrt{s_{NN}}$ = 27, 39, and 62.4~GeV taken
with the STAR detector at the Relativistic Heavy Ion Collider. For all energies studied, a 
significant excess yield of dielectrons is observed in the low-mass region
(0.40$ < M_{ee} < 0.75$ MeV/$c^2$) compared to hadronic cocktail simulations at freeze-out.
Models that include an in-medium broadening of the $\rho$-meson spectral function consistently 
describe the observed excess. In addition, we report acceptance-corrected dielectron-excess spectra 
for Au+Au collisions at mid-rapidity ($\left|y_{ee}\right|$ $<$ 1) in the 0$-$80\% centrality
bin for each collision energy. The integrated excess yields for $0.4 < M_{ee} < 0.75\ \textrm{GeV}/c^{2}$,
normalized by the charged particle multiplicity at mid-rapidity, are compared with previously 
published measurements for Au+Au at $\sqrt{s_{NN}}$ = 19.6 and 200 GeV. The normalized excess yields 
in the low-mass region show no significant collision energy dependence. The data, however, are 
consistent with model calculations that demonstrate a modest energy dependence. 
\end{abstract}

\maketitle

Experimentally, dileptons are good probes of the hot quantum chromodynamics (QCD) medium created in 
heavy-ion collisions because leptons are not affected by the strong interaction. As a result, leptons can 
traverse the hot medium with minimal final-state effects, providing means to experimentally test models 
that predict chiral symmetry restoration, and enabling a better understanding of the microscopic properties 
of QCD matter.

The generation of hadronic masses is in part caused by the spontaneous breaking of chiral symmetry 
\cite{DSE_AspectsOfHadronPhysics,LAtticeReview_FLAGColl}.  Ultrarelativistic heavy-ion collisions produce a 
hot and dense QCD medium, a Quark-Gluon Plasma (QGP), where partial chiral symmetry restoration is expected 
\cite{ChiralLAttice3}.  Theoretical calculations suggest that chiral symmetry restoration will result in 
the modification of chiral partners such as the $\rho(770)$ vector meson and the $a_{1}(1260)$ axial-vector 
meson \cite{RappChiral} with subsequent $\rho$ and $a_{1}$ mass degeneracy.  Reconstruction of the $a_{1}$ 
is an experimentally challenging task with its broad resonance width and decay daughter(s) (e.g., $\pi$) 
that rescatter in the QCD medium.  The $\rho$ however, can be reconstructed through its leptonic 
$e^{+}e^{-}$ decay channel, allowing its spectral distribution to be studied.

The CERES Collaboration at the Super Proton Synchrotron observed an excess yield in Pb$+$Au collisions at 
$\sqrt{s_{NN}}$ = 8 and 17.3~GeV \cite{CERES2003,CERES2000} in the low dielectron (unlike-sign pairs unless 
otherwise specified) invariant mass range (LMR) (i.e., below the $\phi$ meson mass), where excess yield is 
the difference between the measured yield and an expected yield based on simulations.  The excess yield in 
the LMR was observed relative to known hadronic sources, including the $\rho-$decay in vacuum. 
High-precision dimuon measurements in In$+$In collisions by the NA60 Collaboration at
$\sqrt{s_{NN}}$ = 17.3~GeV suggest that the observed LMR excess is consistent with the in-medium broadening 
of the $\rho$ spectral function \cite{NA60}. At the Relativistic Heavy Ion Collider (RHIC), measurements of 
dielectron mass spectra in Au+Au collisions at $\sqrt{s_{NN}}$ = 200~GeV show a significant excess in the 
LMR when compared to the known hadronic sources.  The excess has been observed by both the STAR and PHENIX 
Collaborations \cite{STARAuAu200PRL,STARAuAu200PRC,STARAuAu19200PLB,PHENIXAuAu200PRC81,PHENIX200PRCnew}.  
Theoretical calculations using a many-body approach \cite{Rapp4STAR1,*Rapp4STAR2,*Rapp4STAR3,*Rapp4STAR4,*Rapp4STAR5}
or a transport model \cite{PHSD,*PHSD2,*PHSD_Linnyk,PHSD_dileptonReview} predict an in-medium broadened 
$\rho$ spectral function and both calculations are consistent with the previously published STAR 
measurements at $\sqrt{s_{NN}} = 200$~GeV. 

The RHIC Beam Energy Scan (BES) program \cite{STARBESIproposal} provides a unique opportunity to 
systematically test these calculations as a function of the initial collision energy. In the BES energy 
range between $\sqrt{s_{NN}}$ = 27 and 62.4~GeV we observe the freeze-out temperature \footnote{The 
temperature of the expanding QCD matter when nuclear scatterings cease.} to remain constant 
\cite{STARBulkProperties}. Moreover, we find the total baryon density to remain approximately constant,
based on the yield ratio of protons and anti-protons to charged pions \cite{STARBulkProperties,pik62prc}. 
Both will serve here as a baseline to test the aforementioned theoretical calculations against.

In this Letter, the STAR Collaboration presents the first measurements of dielectron production in Au+Au 
collisions with colliding nucleon + nucleon pair energy ($\sqrt{s_{NN}}$) at 27, 39, and 62.4~GeV. The data 
were collected by the STAR detector in the 2010 and 2011 RHIC runs using a minimum-bias trigger which 
requires a coincidence of signals in the -$z$ and +$z$ components of either the vertex position detector, 
beam-beam counters, or the zero degree calorimeters. The analyses included 68M, 132M, and 62M collision 
events for $\sqrt{s_{NN}}$ = 27~GeV, 39~GeV, and 62.4~GeV, respectively. The main detector systems involved 
in this analysis are the Time Projection Chamber (TPC) \cite{TPC} and the Time-Of-Flight (TOF) detectors 
\cite{TOF}.  We report the LMR acceptance-corrected excess yields for the 80\% most-central collisions.

Electron identification was performed using methods described in \cite{STARAuAu200PRC}.  The TPC is used 
for electron identification via energy-loss measurements and in conjunction with the TOF, the electron 
signal is improved by the removal of slow hadrons. The purity of the electron samples is 95\% for 62.4~GeV 
and 94\% for the other two energies.  The invariant mass spectrum for dielectrons was generated using all 
accepted, oppositely-charged electron candidate pairs from the same event, and summing over all events.  
Only electrons with pseudo-rapidity $\left|\eta^{e}\right|$ $<$ 1 and transverse momentum $p_{T}^{e}$ $>$ 
0.2~GeV/$c$ were used in this analysis.  The dielectrons from photon conversion in the detector materials 
were greatly suppressed by requiring a minimum pair opening angle, as described in 
\cite{PHENIXAuAu200PRC81,STARAuAu200PRC}. The like-sign combination method was adopted to reproduce the 
background because it simultaneously reproduces correlated and uncorrelated sources 
\cite{PHENIXAuAu200PRC81}. The background subtraction was performed as a function of $M_{ee}$ and pair 
momentum $p_{T}^{ee}$.

The raw data were corrected for the single-electron reconstruction efficiency as well as for the
loss of dielectrons in the very low-mass region $M_{ee}$ $<$ 0.2~GeV/c$^{2}$ caused by the minimum opening 
angle requirement. An embedding technique was used to determine the tracking efficiency 
\cite{STARAuAu200PRC,STARAuAu19200PLB}, while the electron identification efficiency was derived from data-
driven techniques \cite{STARAuAu200PRC}. The single-electron reconstruction efficiency was folded into the 
pair efficiency via a virtual photon method \cite{STARAuAu19200PLB} and applied to the background-
subtracted dielectron spectrum in $M_{ee}$ and $p_{T}^{ee}$.

The systematic uncertainties in the final mass spectra include uncertainties in ($i$) the acceptances for 
like-sign and unlike-sign dielectrons, ($ii$) the hadron contamination, and ($iii$) the efficiency 
corrections \cite{STARAuAu200PRC}.  The dominant systematic uncertainty contribution in the LMR, the 
efficiency correction uncertainty is 8\%, 7.7\%, and 10.8\% for $\sqrt{s_{NN}}$ = 27, 39, and 62.4 GeV, 
respectively.  For masses greater than the $\phi$ mass, the hadron contamination uncertainty is of the same 
order as the efficiency-corrections uncertainty.  The acceptance factor uncertainty begins to contribute 
significantly at about 2~GeV/c$^{2}$. The sources of uncertainty are added together in quadrature to 
determine the total systematic uncertainty as a function of $M_{ee}$.

The hadronic sources for dielectrons were simulated using the method described in
\cite{STARAuAu200PRC,STARAuAu19200PLB}, where the meson yields follow the method in 
\cite{STARAuAu19200PLB}. They include contributions from direct and/or Dalitz decays of $\pi^{0}$, $\eta$, 
$\eta'$, $\omega$, $\phi$, $J/\psi$ mesons, as well as contributions from $c\bar{c}$ and Drell-Yan (DY) 
decays.The input $p_T$ spectral shapes were created using Tsallis Blast-Wave (TBW) parameterizations 
\cite{TBWZebo} based on STAR measurements of light hadron production. The J/$\psi$ $p_T$ spectra were 
estimated for $\sqrt{s_{NN}}$ = 39 and 62.4~GeV using Boltzmann parameterizations which were based on 
published data \cite{JPsiWangmeietal}, while the $\sqrt{s_{NN}}$ = 27~GeV spectra were estimated using the 
same parameterization as for the $\sqrt{s_{NN}}$ = 39~GeV data.

The semi-leptonic decays of charmed hadrons in p+p collisions were simulated using PYTHIA v6.416 
\cite{PYTHIA} with the tune described in \cite{HeavyFlavorTune}. The perturbative QCD fixed-order plus 
next-to-leading logarithms upper-limit \cite{CERES2000,FONLL} was used to fit the world-wide measurements 
of $\sigma_{c\bar{c}}^{NN}$ \cite{Charm1,*Charm2,*Charm3,*Charm4} in order to determine the input charm 
production cross-section. The $\sigma_{c\bar{c}}^{NN}$ values estimated from the fit are $26\pm8$~$\mu$b, 
$58\pm16$~$\mu$b, and $130\pm40$~$\mu$b for $\sqrt{s_{NN}}$ = 27, 39, and 62.4~GeV, respectively. The 
obtained charm-related distribution was scaled by the number of nucleon-nucleon binary collisions $N_{\rm 
bin}$ \cite{Glauber} to obtain an estimate of the charm contribution in minimum-bias (0$-$80\% centrality) 
Au$+$Au collisions.  The DY contribution was estimated by following the procedure used in 
\cite{STARAuAu200PRC}. However, $\sigma^{pp}_{DY}$($\sqrt{s}$) was taken from PYTHIA and was corrected by 
the ratio of the cross section used in \cite{STARAuAu19200PLB} to the corresponding PYTHIA cross-section at 
$\sqrt{s}$ = $19.6\textrm{ GeV}$.

\begin{figure}
\begin{centering}
\includegraphics[width=0.49\textwidth]{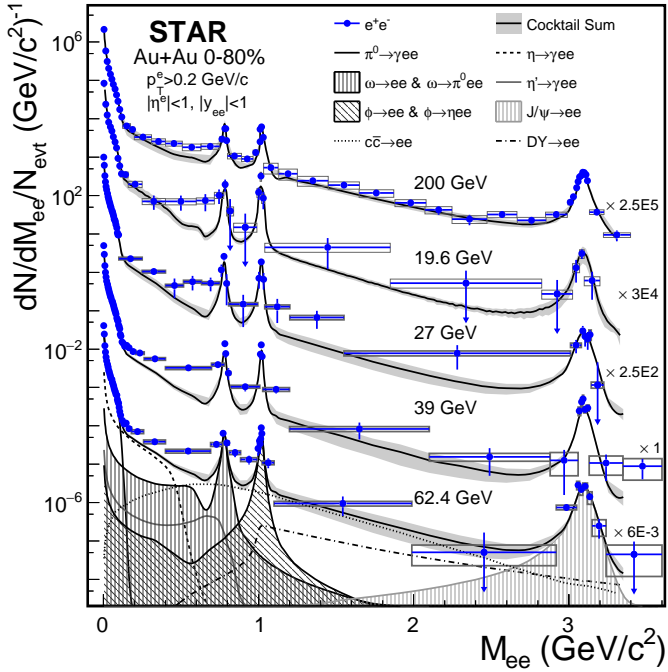}
\par\end{centering}
\protect\caption{ Background subtracted dielectron invariant mass spectra within the STAR acceptance from
$\sqrt{s_{NN}}$ = 19.6, 27, 39, 62.4, and 200~GeV 0$-$80\% most-central Au$+$Au collisions.  Errors bars 
and open boxes represent the statistical and systematic uncertainties in the measurements.  The black solid 
curves represent the hadronic cocktail with the gray bands representing the cocktail uncertainties.  The 
curves underneath the $\sqrt{s_{NN}}$ = 62.4~GeV hadronic cocktail curve and gray band represent the 
cocktail components at 62.4~GeV.  For better presentation, the measurements and cocktail predictions are 
not listed in order by energy but have been scaled by factors $2.5\times 10^5$, $3\times 10^4$, 
$2.5\times 10^2$, 1, and $6\times 10^{-3}$ for results at $\sqrt{s_{NN}}$ = 200, 19.6, 27, 39,
and 62.4~GeV, respectively.}
\label{fig:spectraAll}
\end{figure}

The efficiency-corrected spectra are shown in Fig.\ \ref{fig:spectraAll} for 0$-$80\% most-central Au+Au 
collisions at $\sqrt{s_{NN}}$ = 27, 39, and 62.4~GeV. The figure shows $p_{T}$-integrated invariant mass 
spectra captured in the STAR acceptance at mid-rapidity ($\left|\eta^{e}\right|<1$, $p^{e}_{T}$ $>$ 
0.2~GeV/$c$, and $\left|y_{ee}\right| < 1$), where each data point is positioned at the bin center and the 
bin markers parallel to the x-axis indicate the bin width.  The data are compared to a hadronic cocktail 
without the vacuum $\rho$-meson since its contributions are expected to be strongly modified in the medium.
To illustrate the extent of STAR's systematic study of $e^{+}e^{-}$ production, Fig.\ \ref{fig:spectraAll} 
includes the efficiency-corrected spectra for the 0$-$80\% most-central Au+Au collisions at
$\sqrt{s_{NN}}$ = 19.6 and 200~GeV from Refs.\ \cite{STARAuAu19200PLB,STARAuAu200PRC}.

\begin{figure}
\begin{centering}
\includegraphics[width=0.49\textwidth]{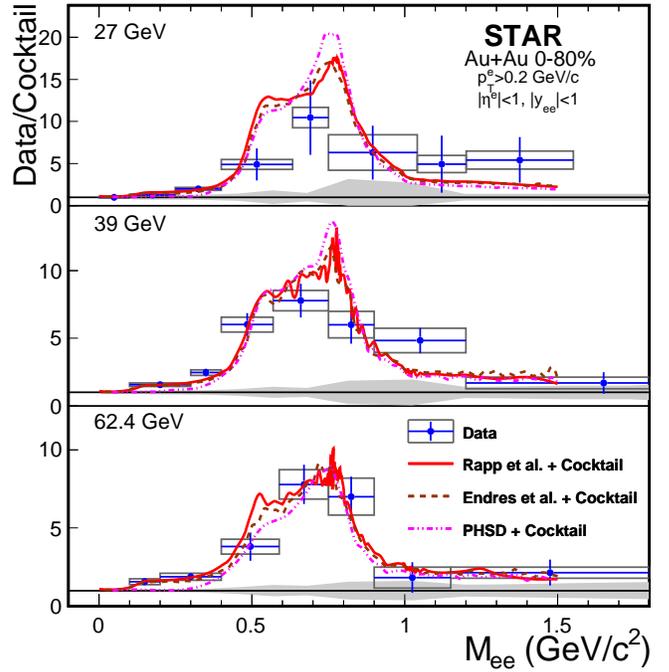}
\par\end{centering}
\protect\caption{ The ratio of the invariant mass spectra to the cocktail with the $\omega$ and $\phi$ 
yields removed from both the data and cocktail. The gray area shows the cocktail uncertainties. Model 
calculations by Rapp \emph{et al.}\ \cite{Rapp4STAR1,*Rapp4STAR2,*Rapp4STAR3,*Rapp4STAR4,*Rapp4STAR5}, 
Endres \emph{et al.}\ \cite{Endres_NA60,*Endres_19200}, and PHSD 
\cite{PHSD,*PHSD2,*PHSD_Linnyk,PHSD_dileptonReview} were separately added to the reference cocktail and compared to the reference cocktail, via ratios, as shown with the curves.}
\label{fig:Data2Cocktail}
\end{figure}

Figure \ref{fig:Data2Cocktail} shows the ratio of the present data to the hadronic cocktail with the yields 
from $\omega$ and $\phi$ subtracted from both the data and cocktail.  The open boxes depict the 
experimental systematic uncertainties, while the gray bands represent the cocktail simulation 
uncertainties.  To keep the data and cocktail uncertainties separate throughout this study, the $\omega$ 
and $\phi$ yield uncertainties remain in the cocktail uncertainties.  A clear enhancement is observed in 
the LMR relative to the hadronic cocktail for each of the three collision energies.

Model calculations within the STAR acceptance by Rapp \emph{et al.\ }\cite{Rapp4STAR1,*Rapp4STAR2,*Rapp4STAR3,*Rapp4STAR4,*Rapp4STAR5,RappHvHLifetime,*RappHvHLifetime2},
Endres \emph{et al.\ }\cite{Endres_NA60,*Endres_19200}, and calculations using the PHSD model
\cite{PHSD,*PHSD2,*PHSD_Linnyk,PHSD_dileptonReview} were separately added to the hadronic cocktail and the 
resulting combined spectra are compared to the reference cocktail, via ratios, as shown in 
Fig.\ \ref{fig:Data2Cocktail}. The model by Rapp \emph{et al.\ }is an effective many-body calculation for 
vector mesons in a QGP where the spectral function of $\rho$ is modified (broadened) primarily due to 
interactions with baryons and mesons (i.e., a hadron gas).  The model by Endres \emph{et al.\ }is a 
coarse-grained transport approach that includes the $\rho$ spectral function mentioned above from 
\cite{WambachRapp,Rapp4STAR1,*Rapp4STAR2,*Rapp4STAR3,*Rapp4STAR4,*Rapp4STAR5}. PHSD is a microscopic 
transport model which includes the collisional broadening of the $\rho$.  Each model has successfully 
described the LMR $\mu^{+}\mu^{-}$ excess yield observed by the NA60 experiment, as well as measurements of 
Au$+$Au collisions at $\sqrt{s_{NN}} = 200\ \textrm{GeV}$ 
\cite{STARAuAu200PRL,PHENIX200PRCnew,Endres_NA60,*Endres_19200,Rapp4STAR1,*Rapp4STAR2,*Rapp4STAR3,*Rapp4STAR4,*Rapp4STAR5,PHSD_dileptonReview}. 
Each model includes thermal contributions from the in-medium broadening of the $\rho$ spectral function and 
a QGP.  In contrast to the models in 
\cite{Rapp4STAR1,*Rapp4STAR2,*Rapp4STAR3,*Rapp4STAR4,*Rapp4STAR5,RappHvHLifetime,*RappHvHLifetime2, Endres_NA60,*Endres_19200}, 
the PHSD model includes an incoherent sum of contributions from the $\rho$, the QGP, and Dalitz decays of 
the $a_{1}$ and $\Delta$ resonances.  These contributions tend to underestimate the $e^{+}e^{-}$ yield for 
$M_{ee} < 0.3 \textrm{ GeV}/c^{2}$.  However, we note that these PHSD model calculations do not include 
Bremsstrahlung processes \cite{PHSD_dileptonReview}. 

\begin{figure}
\begin{centering}
\includegraphics[width=0.49\textwidth]{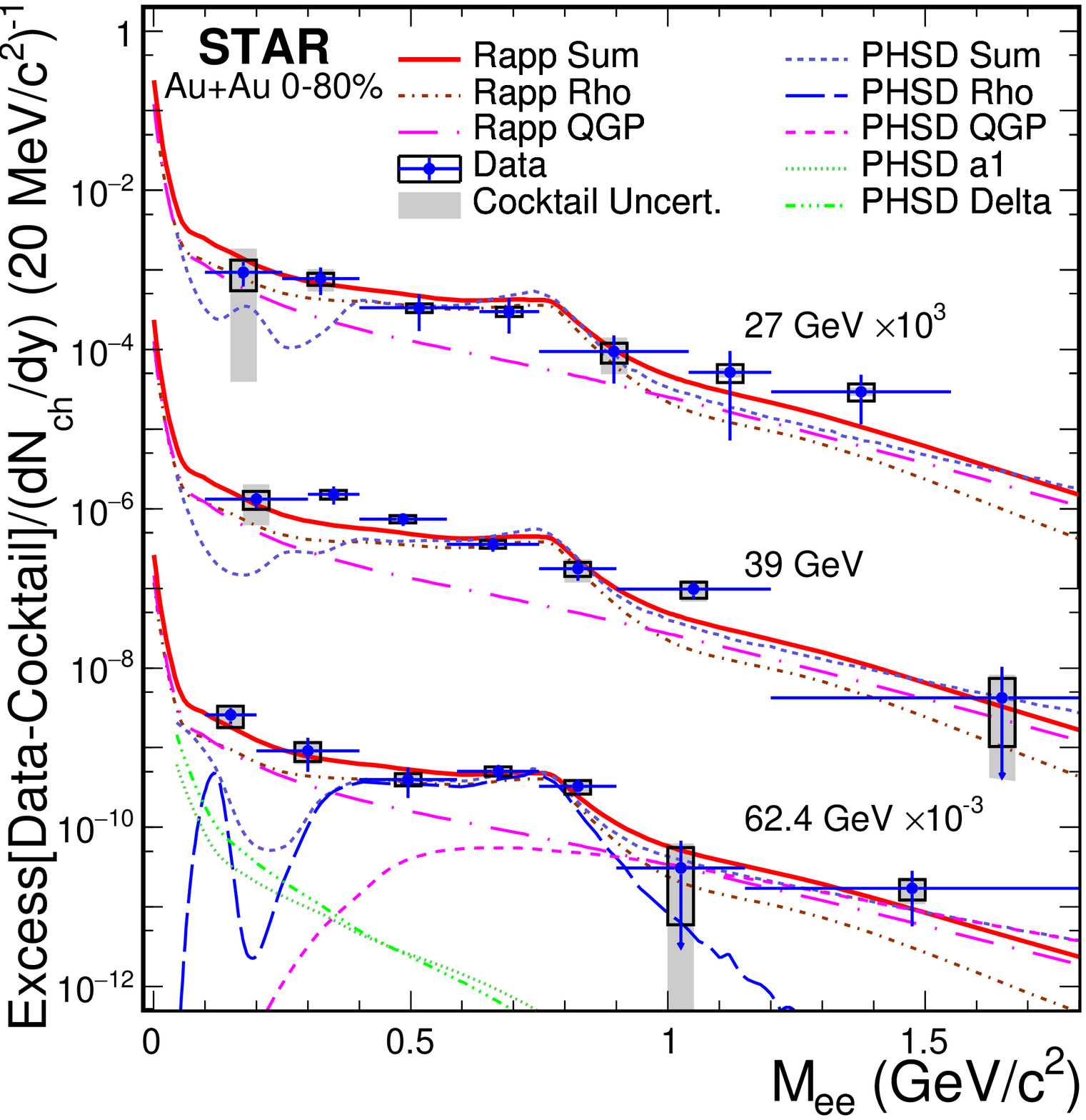}
\par\end{centering}
\protect\caption{ Acceptance-corrected dielectron excess mass spectra, normalized by $dN_{ch}/dy$, for
Au+Au collisions at $\sqrt{s_{NN}}$ = 27, 39, and 62.4~GeV.  Model calculations (curves)
\cite{Rapp4STAR1,*Rapp4STAR2,*Rapp4STAR3,*Rapp4STAR4,*Rapp4STAR5,PHSD,*PHSD2,*PHSD_Linnyk,PHSD_dileptonReview}
are compared with the excess spectra for each energy as explained in the text.  Individual components of 
the PHSD model calculations are only shown for Au+Au collisions at $\sqrt{s_{NN}}$ = 62.4~GeV. The error 
bars, open boxes, and filled boxes indicate statistical, systematic, and cocktail uncertainties.  A 6\% 
uncertainty on the acceptance correction is not shown.}
\label{fig:ExcessVsTheory}
\end{figure}

To further quantify the excess in the LMR, cocktail contributions excluding the $\rho$-meson were 
subtracted from the dielectron yields.  The excess spectra were corrected for the STAR acceptance using a 
virtual photon method similar to that described in \cite{STARAuAu19200PLB}.  The corrected excess yields 
were then normalized to the charged particle multiplicities at mid-rapidity ($dN_{ch}/dy$ \footnote{For 
Au$+$Au collisions at $\sqrt{s_{NN}}$ = 27 and 39~GeV, $dN_{ch}/dy$ is approximated by the $dN/dy$ sum of 
$\pi^{\pm}$, $K^{\pm}$, $p$, and $\bar{p}$ \cite{STARBulkProperties}. For $\sqrt{s_{NN}}$= 62.4~GeV, 
$dN_{ch}/dy$ is given in \cite{pik62prc}}) in order to cancel out the volume effect.
Figure \ref{fig:ExcessVsTheory} shows the acceptance-corrected excess spectra. Systematic uncertainties 
from the measurements and the cocktail are shown in the figure as the open and filled boxes, respectively. 
The 6\% uncertainty from STAR's acceptance correction and the uncertainty of $dN_{ch}/dy$ are not shown in 
the figure.  Model calculations
\cite{Rapp4STAR1,*Rapp4STAR2,*Rapp4STAR3,*Rapp4STAR4,*Rapp4STAR5,PHSD,*PHSD2,*PHSD_Linnyk,PHSD_dileptonReview}
in Fig.\ \ref{fig:ExcessVsTheory} include contributions from broadening of the $\rho$ spectral function in 
a hadron gas (Rapp Rho), and from QGP radiation (Rapp QGP). The PHSD model calculations in
Fig.\ \ref{fig:ExcessVsTheory}  include contributions from the $\rho$ meson (PHSD Rho), QGP (PHSD QGP), 
Dalitz decays of the a$_{1}$ (PHSD a1), and $\Delta$ resonances (PHSD Delta). The sums (Rapp Sum, PHSD Sum) 
are compared with the excess yield at each energy. Calculations by Rapp \emph{et al}.\ have an uncertainty 
on the order of 15\% \cite{Rapp4STAR1,*Rapp4STAR2,*Rapp4STAR3,*Rapp4STAR4,*Rapp4STAR5}, and PHSD model 
calculations have an uncertainty on the order of 30\% \cite{PHSDuncertainty}. Within uncertainties, the 
model calculations are found to reproduce the acceptance-corrected excess in Au+Au collisions at each of 
the collision energies.

\begin{figure}
\begin{centering}
\includegraphics[width=0.425\textwidth]{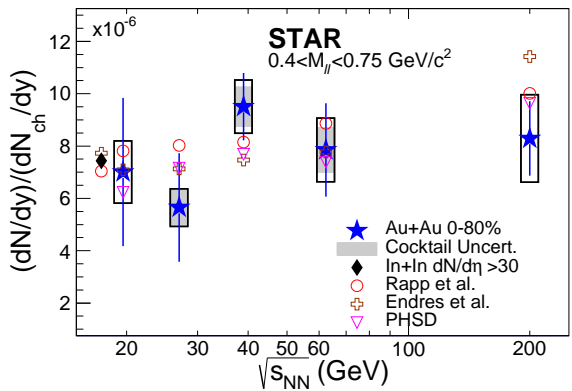}
\par\end{centering}
\protect\caption{Collision energy dependence of the integrated dilepton excess yields in 
$0.4 < M_{ll}< 0.75\ \textrm{GeV}/c^{2}$, normalized by $dN_{ch}/dy$. The closed markers represent the 
experimental measurements while the open markers represent the calculations from Rapp \emph{et al.}, 
Endres \emph{et al.}, and PHSD.  For measurements at $\sqrt{s_{NN}}$ = 27, 39, and 62.4~GeV, the open and 
filled (gray) boxes represent the systematic errors in the measurements and the cocktail uncertainties, 
respectively.  The 6\% uncertainty from the acceptance correction is not included.  For measurements of 
minimum-bias, 0$-$80\% central Au$+$Au collisions at $\sqrt{s_{NN}}$ = 19.6 and 200~GeV, the open boxes 
represent the total systematic uncertainty in the measurements.}
\label{fig:excessVsEnergy}
\end{figure}

To allow for a direct comparison of our measurements with previously published results and model 
calculations, we integrated the acceptance-corrected dielectron excess spectra in the mass region from 
$0.40\textrm{ to }0.75\  \textrm{GeV}/c^{2}$. Figure \ref{fig:excessVsEnergy} shows the integrated excess 
yields normalized by $dN_{ch}/dy$ from the 0$-$80\% most-central Au$+$Au collisions at $\sqrt{s_{NN}}$ = 
27, 39, and 62.4~GeV, together with our previously published results \cite{STARAuAu19200PLB} for the
0$-$80\% most-central Au+Au collisions at $\sqrt{s_{NN}} = 19.6$ and 200~GeV. In addition, we compare to 
the NA60 $\mu^{+}\mu^{-}$ measurement at $\sqrt{s_{NN}}=17.3$\ GeV for $dN_{ch}/d\eta>30$ 
\cite{NA60_Updated}\footnote{{NA60} measurements in \cite{NA60} have been updated in \cite{NA60_Updated}. 
This paper uses the updated measurements while \cite{STARAuAu19200PLB} used the previous measurements. 
Additionally, $dN_{ch}/dy = 120$ is used, where $dN_{ch}/dy = 140$ was used in \cite{STARAuAu19200PLB}}.  
For the measurements at $\sqrt{s_{NN}}$ = 27, 39, and 62.4~GeV, the systematic uncertainties from the data 
and cocktail are shown as the open and filled boxes, respectively.  For the measurements at 
$\sqrt{s_{NN}}$ =  19.6 and 200~GeV, the total (cocktail+data) systematic uncertainties are shown as the 
open boxes.  The normalized, integral yields from model calculations, shown in 
Fig.\ \ref{fig:excessVsEnergy}, are in agreement with the measurements. Note that the result for Au+Au at 
$\sqrt{s_{NN}} = 19.6\ \textrm{GeV}$ \cite{STARAuAu19200PLB} is consistent within uncertainties with the 
$\mu^{+}\mu^{-}$ measurement from NA60 in In$+$In collision at $\sqrt{s_{NN}} = 17.3~\textrm{GeV}$ 
\cite{NA60,NA60_Updated,Note3}. 
 
The normalized integrated excess yields show no statistically significant collision-energy dependence for 
the 0-80\% most-central Au$+$Au collisions.  This may be because dilepton production in the medium is 
expected to be mainly determined by the strong coupling of the $\rho$-meson to baryons, rather than to 
mesons \cite{RappChiral}.  We know that the total baryon density remains approximately unchanged for 
minimum-bias Au$+$Au collisions with collision energies above $\sqrt{s_{NN}}$ = 20~GeV \cite{STARBulkProperties}. 
However, the models and our data are statistically consistent even though the model predictions display 
modest energy dependence.  

In summary, we have reported dielectron yields for the 0$-$80\% most-central Au$+$Au collisions at 
$\sqrt{s_{NN}}$ = 27, 39, and 62.4~GeV.  The data were collected with the STAR detector at RHIC.  The new 
measurements complement the previously published results 
\cite{STARAuAu200PRL,STARAuAu200PRC,STARAuAu19200PLB,PHENIX200PRCnew} and the combined datasets now cover 
an order-of-magnitude range in collision energies over which the total baryon density and freeze-out 
temperatures are remarkably constant \cite{STARBulkProperties}.
Across the collision energies, we have observed statistically significant excesses in the LMR when 
comparing the data to hadronic cocktails that do not include vacuum $\rho$ decay contributions.  The excess 
yields have been corrected for acceptance, normalized by $dN_{ch}/dy$, integrated from 0.40 to 
0.75~GeV/$c^{2}$, and reported as a function of $\sqrt{s_{NN}}$.  The measured yields show no significant 
energy dependence, and are statistically consistent with model calculations. 

Our findings, while restricted to the $\rho$-meson mass range and limited by statistical and systematic 
uncertainties, are consistent with models that include $\rho$ broadening in the approach to chiral symmetry 
restoration \cite{HohlerRapp}.  Further experimental tests of the models discussed in this paper are 
warranted.

As part of the Beam Energy Scan Phase II project, the STAR Collaboration plans to collect over an order-of-
magnitude more data than previously acquired in the energy range from 7.7 to 19.6~GeV, where the total 
baryon density changes substantially \cite{STARBulkProperties}. Future studies may therefore allow us to 
better understand the competing factors that play a role in the LMR dielectron excess production 
\cite{RappHvHLifetime,RappHvHLifetime2} and to further clarify the connection between $\rho$-meson 
broadening and chiral symmetry restoration.  

\begin{acknowledgments}
We thank the RHIC Operations Group and RCF at BNL, the NERSC Center at LBNL, and the Open Science Grid 
consortium for providing resources and support. We also thank S.~Endres, O.~Linnyk, E.~L.~Bratkovskaya, and 
R.~Rapp for discussions and model calculations. This work was supported in part by the Office of Nuclear 
Physics within the U.S. DOE Office of Science, the U.S. National Science Foundation, the Ministry of 
Education and Science of the Russian Federation, National Natural Science Foundation of China, Chinese 
Academy of Science, the Ministry of Science and Technology of China and the Chinese Ministry of Education, 
the National Research Foundation of Korea, GA and MSMT of the Czech Republic, Department of Atomic Energy 
and Department of Science and Technology of the Government of India; the National Science Centre of Poland, 
National Research Foundation, the Ministry of Science, Education and Sports of the Republic of Croatia, 
RosAtom of Russia and German Bundesministerium fur Bildung, Wissenschaft, Forschung and Technologie (BMBF) 
and the Helmholtz Association.
\end{acknowledgments}

\bibliography{BES_PRL}

\end{document}